\documentstyle[11pt,IAU207_pasp,twoside,psfig]{article}
\def\edcomment#1{\iffalse\marginpar{\raggedright\sl#1\/}\else\relax\fi}
\reversemarginpar

\begin{document}
\title{The Globular Cluster Systems of
Ellipticals and Spirals}
 \author{Duncan A. Forbes}
\affil{Astrophysics \& Supercomputing, Swinburne University, 
Hawthorn, VIC 3122, Australia}

\begin{abstract}
In this overview of the key properties of globular cluster (GC) 
systems I show that the GCs in elliptical and spiral host 
galaxies have more in common than previously thought.
After contrasting these properties I briefly comment on GC formation.

\end{abstract}

\section{Spatial Distribution}

For the Milky Way (Minniti 1995) and other spiral galaxies (Forbes, 
Brodie \& Larsen 2001) a consenus is emerging that the 
bulk of the metal-rich globular clusters (GCs) 
in spirals are associated with the bulge rather than the disk 
component, and the metal-poor GCs with the halo. 
A similar situation appears to be present in ellipticals with the
red GCs associated with the elliptical 
bulge/spheroid and
the blue GCs with the (hot gas) halo (Forbes, Brodie \& Larsen
2001). 

Globular cluster systems typically reveal a surface density
distribution that is near 
constant close to the galaxy centre and which falls off rapidly like a 
power-law at larger radii. By fitting a King core profile to the GC density 
profile, a system core radius can be defined. Early-type galaxies 
reveal a trend for more massive galaxies to have larger core radii (Forbes 
et al. 1996). The Milky Way GC system, with a core of $\sim$ 1
kpc, seems to obey this trend.

\section{Numbers and Specific Frequency}

For small ellipticals, the number of GCs scales with the 
galaxy luminosity, while the most massive ellipticals appear to have a 
slightly steeper scaling relation (Djorgovski \& Santiago 1992). 
One possible explanation for this comes from the 
work of McLaughlin (1999). 
This model requires
a universal efficiency factor which is defined to be the ratio of
mass in GCs to that of the initial gas mass available to form
those GCs. For most ellipticals the initial gas supply has been
converted into stars and is hence well represented by the current
stellar mass (or luminosity). 
As a consequence these galaxies have a near
constant S$_N$ $\sim$ 3. 
For high luminosity galaxies, McLaughlin argues that the
current stellar mass grossly underestimates the initial gas
supply and so they appear to be rich in GCs 
when they should be considered as poor in field stars. 
However by including the current gas content (as indicated by the
presence of extensive X-ray gas), along with the current stellar
mass, the GC systems of such galaxies can be understood as part of
a continuous trend. 
McLaughlin's relation can 
successfully explain the wide variation in GC number with luminosity. 

Fig. 1 shows the McLaughlin relation with a 
universal efficiency factor of 0.2\%. The data for early-type galaxies 
scatters uniformly about the relation. I have also 
included the location of 
three well-studied 
spiral galaxies (using the bulge luminosity), i.e. M104 (NGC
4594), M31 and the Milky Way 
(data from Forbes et al. 2000 and Larsen et al. 2001). 
{\it The GC systems of the three spirals are consistent 
with the early-type galaxy number--luminosity trend. }

\begin{figure*}
\centerline{\vbox{}\psfig{figure=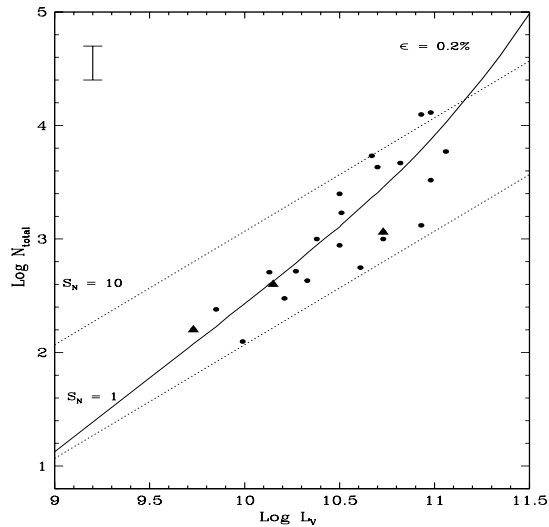,height=7.5cm,width=7.5cm,angle=0}}
\caption{Number of globular clusters versus host galaxy luminosity for
early-type galaxies. 
The spiral galaxies M104, M31 
and the Milky Way are represented by filled triangles, with the
luminosity that of the bulge component only. 
The dashed lines show lines of constant specific frequency, the
solid line shows the universal efficiency model of McLaughlin
(1999) for an efficiency of 0.2\%. The GCs of spirals
follow a similar trend to that of early--type galaxies.  
}
\end{figure*}

\section{Metallicities and Abundances}

Historically, one difference 
between the GCs in spirals and ellipticals was the mean 
metallicity of the two peaks. 
The Milky Way and M31 have GC subpopulations with 
metallicities of [Fe/H] $\sim$ --1.5 and --0.5  
(Barmby et al. 2000) 
while ellipticals were thought to have mean GC [Fe/H] $\sim$ --1.0
and 0.0 (Harris 1991). 
Recently, two developments 
have caused us to reasses the mean GC metallicity 
in ellipticals -- both of which make the GCs in ellipticals more 
metal--poor. 
The first effect is the use of 
more accurate transformations from optical colours
to [Fe/H]. For example, the new 
transformation of 
Kissler--Patig et al. (1998) converts a typical V--I = 1.05 
to [Fe/H] = --1.07, where the old Galactic--based transformation would 
give [Fe/H] $\sim$ --0.5. 
The second effect is that the more accurate 
Galactic extinction values of Schlegel et al. (1998) tend to be larger on 
average by up to A$_V$ $\sim$ 0.1, than the 
traditionally-used Burstein \& 
Heiles (1984) values. Thus extinction-corrected GC colours are now bluer 
than previously, and more metal--poor when transformed. 
If these two effects are taken into account, the two GC subpopulations 
in ellipticals have 
mean metallicities of [Fe/H] $\sim$ --1.5 and --0.5 
which is similar to that
of the M31 and Milky Way GC systems.
{\it Thus there appears
to be very little difference between the mean metallicity of the
two subpopulations in spirals and giant ellipticals. }
 
The GC metallicity distributions for a wide range of Hubble types and 
luminosities is shown in Fig. 2 (including the `Local Group Elliptical'
-- see Forbes et al. 2000). They are remarkably 
similar. All galaxies 
appear to possess a population of metal-poor ([Fe/H] $\sim$ --1.5) and 
presumably old GCs, while all bulge systems also have some metal-rich GCs 
([Fe/H] $\sim$ --0.5). Hilker (this conference) confirms the
bimodality in NGC 3311 with the same peaks. 

\begin{figure*}
\centerline{\vbox{}\psfig{figure=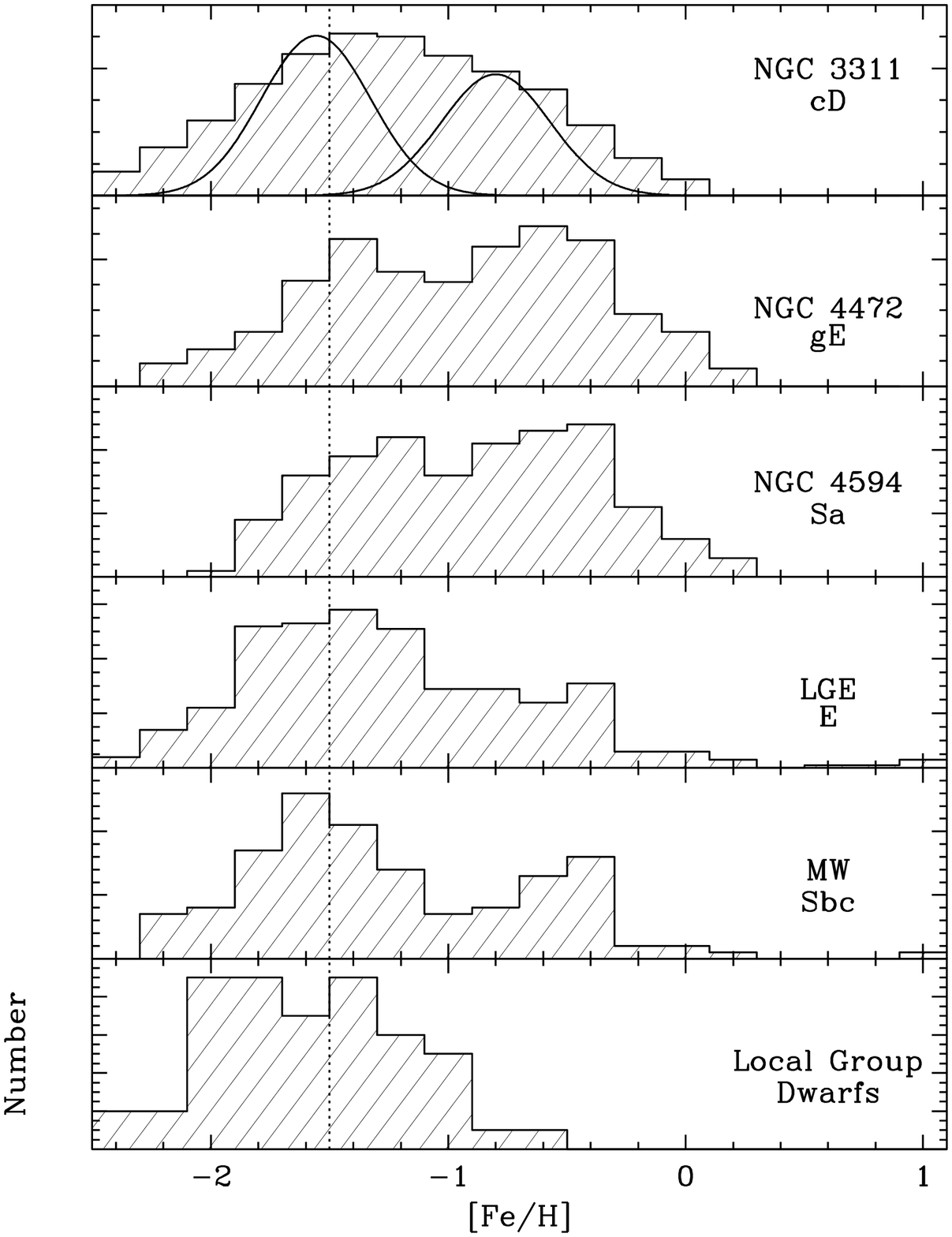,height=7.5cm,width=7.5cm,angle=0}}
\caption{Metallicity distributions for a range of Hubble
types. All galaxies reveal a metal-poor [Fe/H] $\sim$ --1.5 
subpopulation and all
galaxies with bulges reveal a metal-rich [Fe/H] $\sim$ --0.5 
subpopulation.
}
\end{figure*}

There has been some debate over the years about the exact nature
of the GC mean metallicity vs mass relation. 
To investigate this in more detail, 
we have created the largest sample to date of (37) early-type 
galaxies from the literature 
with bimodal GC
colour distributions. This is therefore larger than the Forbes \& 
Forte (2001), Brodie (this conference) or Kundu (this conference) samples.
We
also include M104, M31 and  
the Milky Way. We examine the correlation of colour (metallicity) 
with galaxy velocity dispersion (mass) in Fig. 3.

The Spearman rank correlation test finds that {\it both} the red
and blue GCs are correlated with galaxy velocity dispersion with
probabilities of 99.9\% and 99.2\% respectively.   
A least squares fit gives a positive slope for the red GCs at the
4$\sigma$ level and 2.5$\sigma$ for the blue GCs. 
{\it The mean colors of the GC subpopulations in the three
spirals are consistent with the best fit relation. } 
We also show the galaxy color -- velocity dispersion relation
which has the same slope as for the red GCs. 

\begin{figure*}
\centerline{\vbox{}\psfig{figure=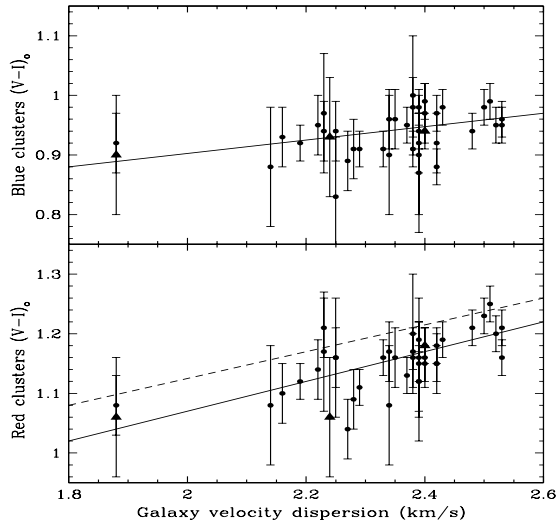,height=7.5cm,width=7.5cm,angle=0}}
\caption{Mean color of globular cluster subpopulations 
versus log galaxy velocity dispersion. Early-type galaxies are shown
by filled circles and spirals by triangles. 
The solid lines show the best fit for the GC colors.
The dashed line shows the galaxy color -- velocity dispersion relation. 
}
\end{figure*}

The existence of a GC color (metallicity) -- galaxy mass relation, 
indicates a common chemical enrichment history for the red 
GCs and the host galaxy (Forbes \& Forte 2000). 
Additionally, this relation has a similar slope to the galaxy
bulge metallicity--mass relation. 
Thus the red GCs and the galaxy (bulge) stars appear to have
formed in the same star formation event with fairly similar
ages and metallicities for both early and late type galaxies. 
The red GCs and bulges of spirals follow these same trends and
hence a similar evolutionary history. 

The situation for the blue GCs is not quite so clear. The slope
and hence statistical significance of a relation between colour and
mass is less than for the red GCs. If the slope is indeed zero,
i.e. there exists a constant mean colour of about (V--I)$_o$ =
0.95, then it indicates that the blue GCs formed before the
potential well of the galaxy did. This suggests a `pregalactic'
origin (Forbes, Brodie \& Grillmair 1997). 
We note that the model of GC formation at cosmological
reionization proposed by Cen (2001) predicts no strong
correlation between blue GC mean colour (metallicity) and galaxy
mass. 
Alternatively, a
relation with galaxy mass would indicate that they formed
{\it in situ} in the current potential well.

In terms of abundances, 
it appears that from the few GC spectra available  
GCs in ellipticals 
have normal, ie Galactic, abundance ratios. This implies [Mg/Fe] $\sim$ 
+0.3. 
These supersolar abundances suggest either a very short star formation 
timescale or an IMF that is skewed to high mass stars. It will be
interesting 
to extend this work to GCs in low luminosity galaxies for which the 
field stars have lower [Mg/Fe] ratios (Terlevich \& Forbes 2001). 
From the limited data available it seems that GCs in spirals 
and ellipticals 
have the same abundance ratios.

\section{Sizes}

An advantage of HST over ground-based telescopes for the study of GCs is the 
ability of HST to measure GC sizes. The largest study to date is that of 
Larsen et al. (2001). In Fig. 4 we show the effective radius for the blue 
and red GC subpopulations in a wide range of galaxies, i.e. from spiral to 
elliptical to cD galaxy. The figure shows that the general trend of the blue 
GCs being $\sim$20\% larger than the red ones exists over this 
large range in 
Hubble types. What ever process is causing this difference (eg 
destruction on 
different orbits or initial conditions) it 
is doing so irrespective of the host 
galaxy. 

\begin{figure*}
\centerline{\vbox{}\psfig{figure=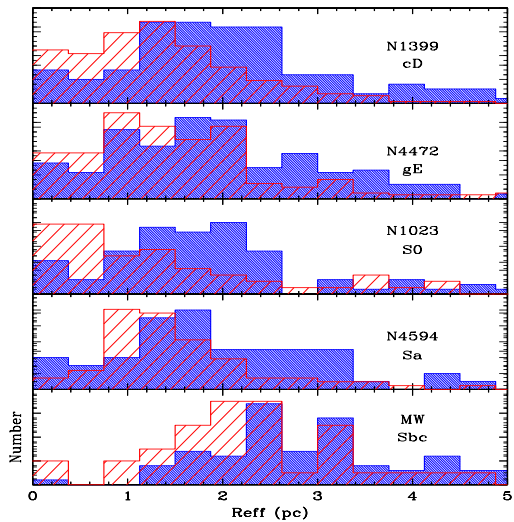,height=7.5cm,width=7.5cm,angle=0}}
\caption{Size of blue and red globular clusters. Blue clusters
(shaded) 
are on average 20\% larger than red clusters (hashed) over a wide range of
host galaxy Hubble type and luminosity. 
}
\end{figure*}

\section{Ages and Kinematics}

Determining the ages, and to a lesser extent kinematics, 
of GCs is perhaps the key to discriminating between 
the different formation models.
Although controversial it appears that the metal-rich 
GCs in our Galaxy may be 
$\sim$ 2 Gyrs younger than the metal-poor ones. Observations to 
date of GCs in ellipticals 
suggest that both subpopulations are old and possibly 
coeval -- a difference 
of a few Gyrs can not be ruled out. 

Kinematics from large numbers of GCs 
are now available for one elliptical (M49) and two cD 
galaxies (M87, NGC 1399). Kinematics also hold important 
clues about formation mechanisms, although we note that both the merger 
and  multi-phase collapse scenarios would probably expect the
blue GCs to show 
higher V/$\sigma$ than the red ones.  See Bridges (this
conference) for further details.

\section{Concluding Remarks}

In terms of the spatial distribution, numbers, chemical
properties, and sizes the GCs in spirals and ellipticals are
remarkably similar. It is perhaps too early to say about the ages
and kinematics. However we should take seriously the possibility
that GC systems formed in a similar way in all galaxies. 
We suggest the following formation timeline. 
About 15 Gyrs ago, GCs with [Fe/H] 
$\sim$ --1.5 formed in all galaxies. They are associated with galaxy halos. 
A few Gyrs later, metal-rich [Fe/H] $\sim$ --0.5 GCs form along with galaxy 
bulges in a clumpy gaseous collapse. 
This is the main epoch of galaxy field star formation (see
Forbes, Brodie \& Grillmair 1997 for more details). In addition, 
recent galaxy 
mergers form elliptical galaxies with moderate S$_N$ values (ie $\sim$ 3).

\section{Acknowledgments}

I thank my SAGES colleagues M. Beasley, 
J. Brodie, C. Grillmair, J. Huchra, M. 
Kissler-Patig, S. Larsen. 
Part of this research was funded by the Ian Potter Foundation 
and NSF grant AST 9900732.


\begin{references}
\reference Djorgovski, S., Santiago, B., 1992, ApJ, 391, L85

\reference Burstein, D., Heiles, C., 1984, ApJS, 54, 33

\reference Cen, R., 2001, astro-ph/0101197

\reference Forbes, D. A., et al. 1996, AJ, 467, 126

\reference Forbes, D. A., Brodie, J. P., Grillmair, C. J.,
1997, AJ, 113, 1652

\reference Forbes, D. A., et al. 2000, A\&A, 358, 471

\reference Forbes, D. A., Forte, J. C., 2001, MNRAS, 322, 257

\reference Forbes, D. A., Brodie, J. P., Larsen, S. S., 2001,
ApJ, submitted

\reference Harris, W. E., 1991, ARA\&A, 29, 543

\reference Kissler-Patig, M., et al. 1998, AJ, 115, 105

\reference Larsen, S. S., Forbes, D. A., Brodie, J. P., 2001,
MNRAS, in press

\reference Larsen, S. S., et al. 2001,
AJ, in press

\reference McLaughlin, D., 1999, AJ, 117, 2398

\reference Minniti, D., 1995, AJ, 109, 1663

\reference Schlegel, D. J., Finkbeiner, D. P., Davis, M.,
1998, ApJ, 500, 525

\reference Terlevich, A. I., Forbes, D. A., 2001, MNRAS, submitted

\end{references}
\end{document}